\documentclass{emulateapj}

\usepackage{graphicx, subfigure}
\usepackage{amssymb}
\usepackage{amsmath}
\usepackage{epstopdf}
 \usepackage{hyperref}
\usepackage{color}
\usepackage{placeins}
\usepackage{natbib}
\usepackage{enumerate}
\usepackage[retainorgcmds]{IEEEtrantools}
\usepackage{textcomp}
\usepackage{footmisc}
\usepackage{xspace}

\newcommand{\Fermi}{\emph{Fermi}\xspace}
\newcommand{\fermi}{\emph{Fermi}\xspace}
\newcommand{\yohkoh}{\emph{Yohkoh}\xspace}
\newcommand{\hsi}{\emph{RHESSI}\xspace}

\newcommand{\soho}{\emph{SOHO}\xspace}
\newcommand{\stereo}{\emph{STEREO}\xspace}
\newcommand{\sdo}{\emph{SDO}\xspace}

\def\Rs{$R_\odot$\xspace}
\def\p0{$\pi^{\rm 0}$}
\def\Angst{$\buildrel _{\circ} \over {\mathrm{A}}$}
\def\de{$^{\circ}$\xspace}
\def\e{\epsilon}
\def\g{\gamma}

\shorttitle{Gamma-rays from a Behind-the-limb Solar Flare}
\shortauthors{\Fermi-LAT Collaboration}

\begin{document}

\title{FIRST DETECTION OF $>100$ MeV GAMMA RAYS ASSOCIATED WITH A
  BEHIND-THE-LIMB SOLAR FLARE}
\author{
  M.~Pesce-Rollins\altaffilmark{1,2,3},
  N.~Omodei\altaffilmark{2,4},
  V.~Petrosian\altaffilmark{2,5},
  Wei~Liu\altaffilmark{2}, 
  Fatima~Rubio~da~Costa\altaffilmark{2},
  A.~Allafort\altaffilmark{2},
  Qingrong~Chen\altaffilmark{2}
}
 
\altaffiltext{1}{Istituto Nazionale di Fisica Nucleare, Sezione di Pisa, I-56127 Pisa, Italy}
\altaffiltext{2}{W. W. Hansen Experimental Physics Laboratory, Kavli Institute for Particle Astrophysics and Cosmology, Department of Physics and SLAC National Accelerator Laboratory, Stanford University, Stanford, CA 94305, USA}
\altaffiltext{3}{email: melissa.pesce.rollins@pi.infn.it}
\altaffiltext{4}{email: nicola.omodei@stanford.edu}
\altaffiltext{5}{email: vahep@stanford.edu}

\begin{abstract}

We report the first detection of $>$100~MeV gamma rays associated with a
behind-the-limb solar flare, which presents a unique opportunity to probe the
underlying physics of high-energy flare emission and particle acceleration. On
2013 October 11 a GOES M1.5 class solar flare occurred $\sim$ 9\de.9 behind the
solar limb as observed by \stereo-B. \hsi observed hard X-ray emission above the
limb, most likely from the flare loop-top, as the footpoints were occulted.
Surprisingly, the \Fermi Large Area Telescope (LAT) detected $>$100 MeV
gamma-rays for $\sim$30 minutes with energies up to 3~GeV. The LAT emission
centroid is consistent with the \hsi hard X-ray source, but its uncertainty does
not constrain the source to be located there.  The gamma-ray spectra can be
adequately described by bremsstrahlung radiation from relativistic electrons
having a relatively hard power-law spectrum with a high-energy exponential
cutoff, or by the decay of pions produced by accelerated protons and ions with
an isotropic pitch-angle distribution and a power-law spectrum with a number
index of $\sim$3.8. We show that high optical depths rule out the gamma rays
originating from the flare site and a high-corona trap model requires very
unusual conditions, so a scenario in which some of the particles accelerated by
the CME shock travel to the visible side of the Sun to produce the observed
gamma rays may be at work.
\end{abstract}


\section{Introduction}
During its first six years in orbit, the \Fermi Large Area
Telescope~\citep[LAT;][]{LATPaper} has detected $>$30 MeV gamma-ray emission
from more than 40 solar flares, nearly 10 times more than
EGRET~\citep{Thompson:93} onboard the {\it Compton Gamma-Ray Observatory},
GRS~\citep{forr85} onboard the {\emph{Solar Maximum Mission} (\emph{SMM})} and
CORONAS-F~\citep{coronasf}. The \Fermi detections sample both the
impulsive~\citep{2012ApJ...745..144A} and the long-duration
phases~\citep{0004-637X-787-1-15} including the longest extended emission ever
detected ($\sim$20 hours) from the SOL2012-03-07 GOES X-class
flares~\citep{0004-637X-789-1-20}.

Our understanding of solar flares has also been shaped by decades of
hard X-ray (HXR) observations, notably by the detection of conjugate footpoint sources
by {\it SMM} \citep{HoyngP1981ApJ246L.155.HXRfp}, coronal sources above 
soft X-ray loops by \yohkoh \citep[e.g.,][]{Masuda1994Nature, 2002ApJ...569..459P},
and double coronal sources suggestive of magnetic reconnection in between
by \hsi \citep[e.g.,][]{SuiL2003ApJ...596L.251S, LiuW_2LT.2008ApJ...676..704L,
LiuW.cusp.flare.20120719M7.2013ApJ...767..168L}. These and many other observations
support the standard flare model involving magnetic reconnection and associated
particle acceleration in the corona \citep[for reviews, see, e.g.,][]{2011SSRv..159..107H}. There are alternatively proposed scenarios, including (re-)accleration of particles in the chromosphere \citep[e.g.,][]{2008ApJ...675.1645F,2012ApJ...749..166H}, supported by \citep[e.g.,][]{2012ApJ...753L..26M}.	

Of particular interest are those flares whose footpoints are occulted by the solar limb, allowing coronal emission to be imaged in greater detail than in normal situations dominated by bright footpoints~\citep[e.g.,][]{KruckerS_Coronal60keV_2007ApJ...669L..49K,KruckerS.occult-stat2008ApJ...673.1181K}.

In this Letter, we present \Fermi and \hsi observations of such a flare whose
position was confirmed to be behind the limb by \stereo-B. While gamma-ray
emission up to tens of MeV resulting from proton interactions has been detected
before from occulted solar flares~\citep[e.g.,][]{1993ApJ...409L..69V,
1994ApJ...425L.109B, 1999A&A...342..575V}, the significance of this particular
event lies in the fact that this is the first detection of $>$100~MeV gamma-ray
emission from a footpoint-occulted flare and presents a unique opportunity to
diagnose the mechanisms of high-energy emission and particle acceleration in
solar flares.

\section{Observations and data analysis}\label{sec:analysis}

\subsection{Observational Overview}
\label{sec:overview}
\begin{figure}[htb!]
\begin{center}
\includegraphics[width=0.49\textwidth]{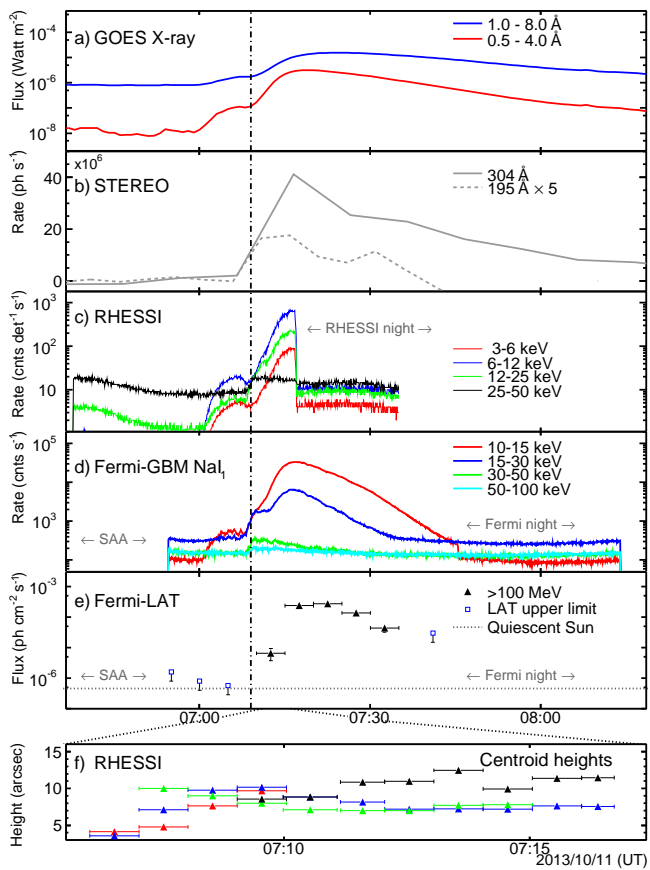}\\
\caption{Light curves of the 2013 October 11 flare as detected by a)
\emph{GOES}, b) \stereo, c) \hsi, d) GBM, e) LAT, and heights of the \hsi
emission centroid (f) with the same color coding as in c). \Fermi exited the
South Atlantic Anomaly (SAA) at 06:57:00 UT. The vertical dashed line represents
flare start time (7:01 UT). } 
\label{fig:LightCurve}
\end{center}
\end{figure}

On 2013 October 11 at 07:01 UT a GOES M1.5 class flare occurred with soft X-ray
emission lasting 44 min and peaking at 07:25:00 UT. Figure~\ref{fig:LightCurve}
shows the GOES, \stereo-B, \hsi, \Fermi Gamma-ray Burst
Monitor~\citep[GBM;][]{meeg09} and LAT lightcurves of this flare. The LAT
detected $>$100~MeV emission for $\sim$30 min with the maximum of the flux
occurring between 07:20:00--07:25:00 UT. \hsi coverage was from
07:08:00--07:16:40 UT, overlapping with \Fermi for 9 min.

Images in Figure~\ref{fig:STEREOSDO} from the \stereo-B Extreme-Ultra Violet
Imager~\citep[EUVI;][]{EUVIinstrument} and the \sdo Atmospheric Imaging
Assembly~\citep[AIA;][]{AIApaper} of the photosphere indicate that the active
region (AR) was $\sim$9\de.9 behind the limb at the time of the flare. LASCO
onboard the \emph{Solar and Heliospheric Observatory} (\soho) observed a
backside asymmetric-halo CME associated with this flare beginning at 07:24:10 UT
with a linear speed of 1200 km s$^{-1}$~\citep{CMEcatalog} and a bright front
over the Northeast. Both \stereo\ spacecrafts detected energetic electrons,
protons, and heavier ions including helium, as well as type-II radio bursts
indicating the presence of a coronal--heliospheric shock. \stereo-B had an
unblocked view of the entire flare and detected a maximum rate of
3.5$\times$10$^{6}$ photons s$^{-1}$ in its 195 \Angst\ channel, corresponding
to a GOES M4.9 class~\citep{2013SoPh..288..241N} if it had not been occulted.

\subsection{Data analysis}\label{sec:analysis}
We performed an unbinned likelihood analysis of the LAT data with the
\texttt{gtlike} program distributed with the \Fermi
\texttt{ScienceTools}\footnote{We used version 09-30-01 available from the
\Fermi Science Support Center \url{http://fermi.gsfc.nasa.gov/ssc/} . We
selected \texttt{P7REP\_SOURCE\_V15} photon events from a 12$^{\circ}$ circular
region centered on the Sun and within 100\de from the local zenith (to reduce
contamination from the Earth’s limb).} For \hsi data, we applied the CLEAN
imaging algorithm~\citep{Hurford2002} using the detectors 3$-$9 to reconstruct
the X-ray images.  We used the FITS World Coordinate System software
package~\citep{2010SoPh..261..215T} to co-register the flare location between
\stereo\ and \sdo images. The \stereo\ light curves are pre-flare background
subtracted, full-Sun integrated photon rates.

\subsection{Localization of the Emission}\label{sec:positions}
\begin{figure}[ht]
\begin{center}
\includegraphics[width=0.48\textwidth]{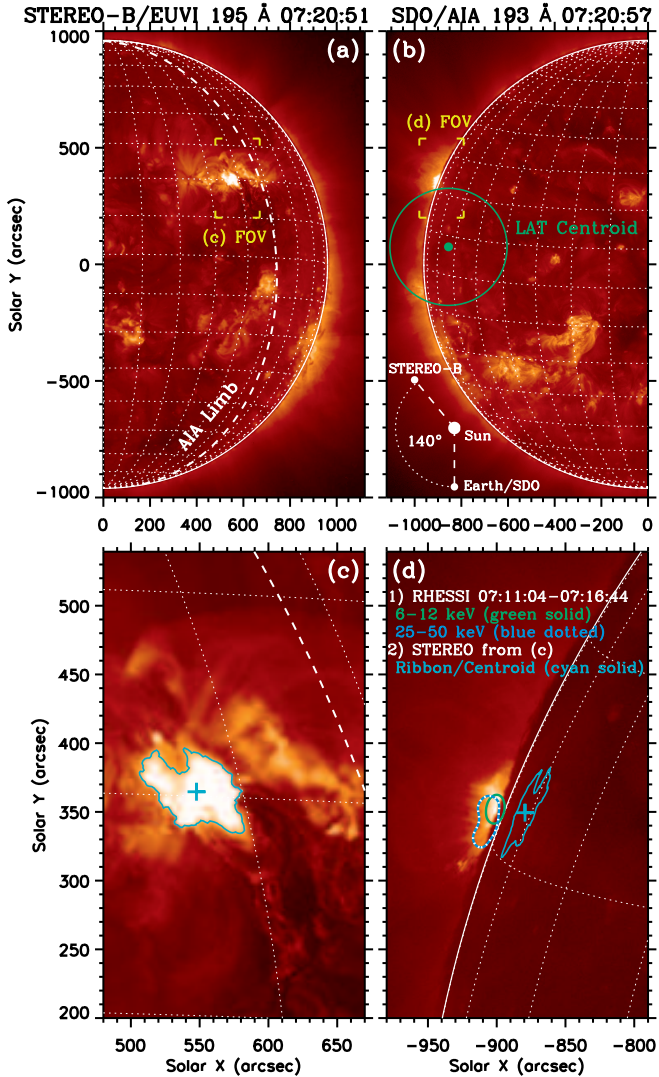}
\caption{\stereo-B (left) and \sdo (right) images near the flare peak. The
white-dashed line in (a) and (c) represents the solar limb as seen by \sdo. The
green line in (b) shows the 68\% error circle for the LAT source centroid. The
cyan contour and plus sign in (c) mark the {\it STEREO} flare ribbon and its
centroid, respectively. Their projected view as seen from the AIA perspective is
shown in (d), in which the centroid is located at 9\de.9 behind the limb. The
green and blue-dotted contours in (d) show \hsi sources. The rectangular
brackets in (a) and (b) mark the field of view (FOV) for (c) and (d),
respectively.}
\label{fig:STEREOSDO}
\end{center}
\end{figure}

We measure the location of the LAT $>$100 MeV gamma-ray emission \citep[as
described in][]{0004-637X-789-1-20} and find a best fit position for the
emission centroid at heliocentric coordinates of ($-850\arcsec$,$70\arcsec$)
with a 68\% error radius of 250$\arcsec$, as shown in
Figure~\ref{fig:STEREOSDO}(b). \hsi\ X-ray sources integrated over
07:11:04$-$07:16:44~UT are shown as 80\%-level, off-limb contours in
Figure~\ref{fig:STEREOSDO}(d). 

The temporal evolution of the projected \hsi source heights above the solar limb
are shown in Figure~\ref{fig:LightCurve}(f). The higher-energy emission
generally comes from greater heights, consistent with expectations for a
loop-top source~\citep[e.g.,][]{SuiL2003ApJ...596L.251S,
2004ApJ...611L..53L}. 
Moreover, from \sdo/AIA
movies we find no signature of EUV ribbons, even in the late phase during the
\hsi night. Together, these observations provide convincing evidence that the
footpoints were indeed occulted. 

The LAT measured 4 photons with energies $>$1 GeV and reconstructed direction
less than 1$^{\circ}$ from the center of the solar disk. All, including a 3~GeV
photon, arrived after 7:19:00 UT (outside of the RHESSI coverage).

\subsection{Spectral analysis}\label{sec:spectra}
We fit the LAT gamma-ray spectral data with three models. The first two, a pure
power-law (PL) and a power-law with an exponential cut-off (PLEXP) are
phenomenological functions that may describe bremsstrahlung emission from
accelerated electrons. The third model uses templates based on a detailed study
of the gamma rays produced from pion decay~\citep[updated from][]{murp87}. 

We rely on the likelihood ratio test~\citep[TS;][]{Mattox:96} to estimate the
significance of the source (TS$_{\rm PL}$) as well as to estimate whether the
addition of the exponential cut-off is statistically significant. To this end we
define $\Delta$TS=TS$_{\rm PLEXP}$-TS$_{\rm PL}$ which is equivalent to the
corresponding difference of maximum likelihoods computed between the two models.
The significance in $\sigma$ can be roughly approximated as $\sqrt{\rm TS}$.

\begin{figure*}[t!]
\begin{center}
\includegraphics[width=\textwidth]{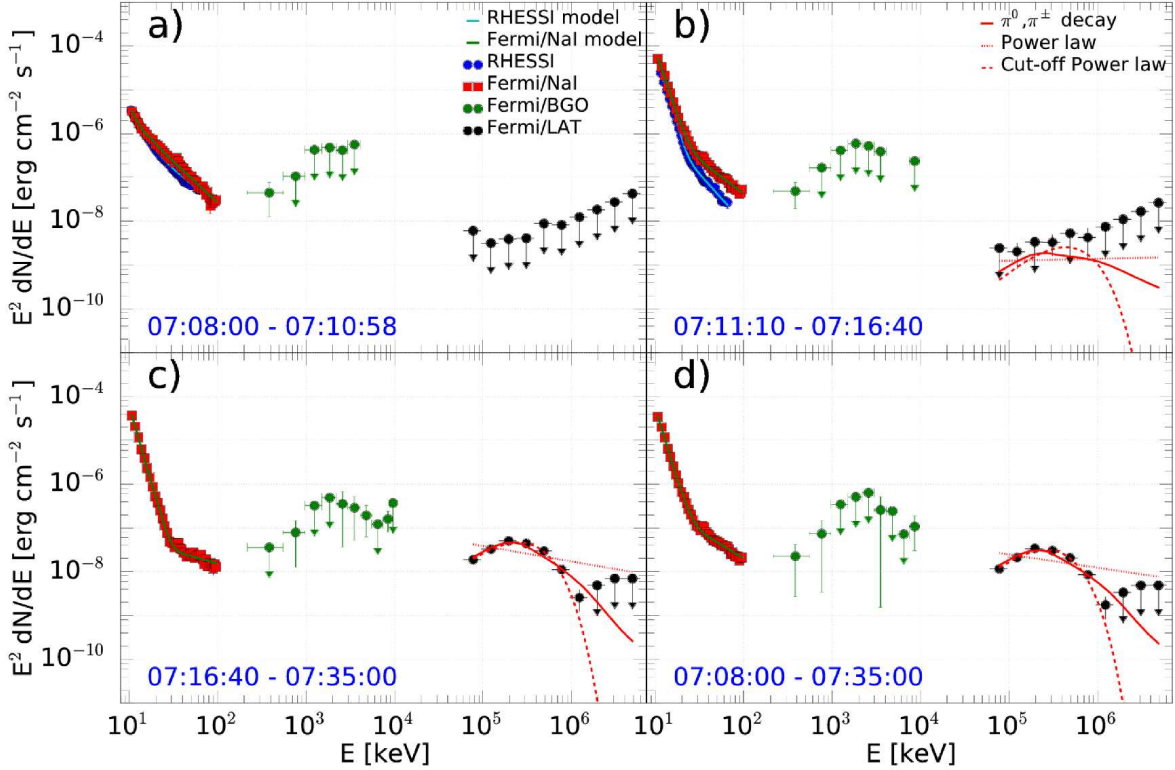}\\
\caption{Spectral energy distributions for \hsi (blue), GBM NaI$_1$ (red),
BGO$_0$ (green) and LAT (black) for four time intervals. Best fit models are
shown in cyan (\hsi), yellow (GBM) and red (LAT).}
\label{fig:SED}
\end{center}
\end{figure*}

For each interval where PLEXP provides a significantly better fit than PL
($\Delta{\rm TS}>$20) we also fit the data with a series of pion-decay models to
determine the best proton spectral index following the same procedure described
in~\citet{0004-637X-789-1-20}. The TS values for PLEXP and pion-decay fits
cannot be directly compared~\citep{wilks1938} however the PLEXP approximates the
shape of the pion-decay spectrum (see Figure~\ref{fig:SED}) thus we expect the
pion-decay models to provide a similarly acceptable fit. We studied the effect
of the LAT systematic uncertainties (mainly from the effective area, as
considered here) via the bracketing technique described
in~\citet{2012ApJS..203....4A}.

The \hsi and GBM NaI$_{1}$ spectral data are independently fitted with one or
two thermal components plus a broken power-law with index fixed to 2 (to avoid
energy divergence) below the break. Table~\ref{tab:LATintervals} summarizes the
spectral analysis results.
Data from BGO$_{0}$ are analyzed using the procedure described in
\citet{2012SPIE.8443E..3BF} with an additional 5\% systematic error on the
background estimation.
Figure~\ref{fig:SED} shows the combined spectra from \hsi, GBM and LAT in four
integration intervals. The discrepancy (up to a factor of 2.5) between the \hsi
and GBM flux values is likely due to pile-up in the \hsi detector, and cannot be
easily corrected. As is evident from Figure~\ref{fig:SED}, more energy is
radiated in HXRs than gamma rays. 

\begin{deluxetable*}{lcccccc}
\tabletypesize{\scriptsize}
\tablecaption{Spectral analysis of LAT, GBM and \hsi data}
\tablewidth{\textwidth}
\tablehead{ 
\multicolumn{7}{c}{LAT time intervals}\\
\hline
\colhead{Time Interval} & \colhead{TS$_{\rm PL}$} & \colhead{$\Delta$TS$^{a}$} &
\colhead{Photon index} & \colhead{Cutoff energy$^{b}$} &  \colhead{Proton index}
& \colhead{Flux$^{c}$} \\
\colhead{(2013/10/11 UT)}  & \colhead{} & \colhead{} & \colhead{} &
\colhead{(MeV)} & \colhead{} & \colhead{($10^{-5}$\,ph cm$^{-2}$\, s$^{-1}$)} }
\startdata
\hline
 07:10:00--07:15:00 & 16 & 5 & 1.80$\pm$0.35$^{+0.02}_{-0.02}$$^{d}$ &--&-- &
0.60$\pm$0.26$^{+0.05}_{-0.04}$ \\ 
 07:15:00--07:20:00 & 987 & 22 & 0.21$\pm$0.34$^{+0.16}_{-0.14}$ &
145$\pm$27$^{+9}_{-8}$& 3.7$\pm$0.2$\pm$0.1 & 24.1$\pm$1.5$^{+1.6}_{-1.4}$ \\ 
 07:20:00--07:25:00 & 1146 & 92 & 0.23$\pm$0.27$^{+0.11}_{-0.11}$ &
162$\pm$26$^{+7}_{-7}$& 3.5$\pm$0.2$\pm$0.1 & 28.2$\pm$1.7$^{+1.7}_{-1.5}$ \\ 
 07:25:00--07:30:00 & 435 & 45 & -0.42$\pm$0.58$^{+0.14}_{-0.15}$ &
95$\pm$21$^{+4}_{-4}$& 4.3$\pm$0.4$\pm$0.1 & 13.7$\pm$1.33$^{+0.8}_{-0.7}$ \\ 
 07:30:00--07:35:00 & 55 & 4& 2.36$\pm$0.24$^{+0.03}_{-0.02}$$^{d}$ &--&--& 
4.1$\pm$1.1$^{+0.4}_{-0.3}$\\ 
 \hline
 07:08:00-- 07:35:00 & 2885 & 233& 0.2$\pm$0.2$^{+0.135}_{-0.132}$ &
147$\pm$15$^{+1}_{-3}$ & 3.7$\pm$0.2$^{+0.1}_{-0.1}$ &
14.2$\pm$0.5$^{+0.9}_{-1.2}$\\
 07:16:40-- 07:35:00 & 2855 & 204& 0.4$\pm$0.2$^{+0.134}_{-0.128}$ &
155$\pm$16$^{+1}_{-2}$ & 3.8$\pm$0.2$^{+0.1}_{-0.1}$ &
22.1$\pm$0.8$^{+1.5}_{-1.8}$ \\
 \hline\\
\multicolumn{7}{c}{\hsi and GBM time intervals}\\
\hline 
& &\multicolumn{2}{c}{Broken power law} & \colhead{1$^{st}$ thermal component} &
\colhead{2$^{nd}$ thermal component}\\
& & \colhead{E$_{break}$}  & \colhead{Index} & \colhead{Plasma temperature}  &
\colhead{Plasma temperature}  & \colhead{Flux$^{e}$} \\
& & \colhead{(keV)} & \colhead{} & \colhead{(keV)}  & \colhead{(keV)}  &
\colhead{(ph cm$^{-2}$ s$^{-1}$)} \\
\hline\hline
07:08:00--07:10:58$^{\star}$ 	& & 17.9$\pm$0.5  & 3.76$\pm$0.04	&
2.23$\pm$0.06 		& -- 				&   52$\pm$1  \\
07:08:00--07:10:58 			& & 17.9$\pm$0.9  & 3.88$\pm$0.03
& 1.9$\pm$0.1 		& 	--			&   55$\pm$1 \\
07:11:10--07:16:40$^{\star}$ 	& & 16$\pm$2	&  4.24$\pm$0.07 	&
1.92$\pm$0.02		& 0.62$\pm$0.03 	& 253$\pm$5  \\
07:11:10--07:16:40  			& & 21$\pm$5	& 3.52$\pm$0.05 
& 2.9$\pm$0.4 			& 1.54$\pm$0.16 	& 630$\pm$10  \\
07:16:40--07:35:00  			& & 20 (fixed)	& 2.56$\pm$0.06	&
2.71$\pm$0.07 			& 1.27$\pm$0.04 	& 399$\pm$8  \\
07:08:00--07:35:00   			& & 20$\pm$8	&  3.22$\pm$0.05
& 2.8$\pm$0.2 			& 1.34$\pm$0.08 	& 388$\pm$8  \\
\enddata
\label{tab:LATintervals}
\tablenotetext{a}{$\Delta$TS=TS$_{\rm PLEXP}$-TS$_{\rm PL}$}
\tablenotetext{b}{From the fit with PLEXP model.}
\tablenotetext{c}{Integrated flux between 100\,MeV and 10\,GeV calculated for
the best fit model.}
\tablenotetext{d}{Photon index from the fit with PL.}
\tablenotetext{e}{Integrated flux between 10 and 100\,keV calculated for the
best fit model.}
\tablenotetext{$\star$}{\hsi spectral fit results. At 07:10:59 UT \hsi changed
its attenuator.}
\tablenotetext{}{Intervals a) and b) of Figure~\ref{fig:SED} excluded due to
lack of statistics. Statistical errors are shown first and
systematic errors follow. }
\end{deluxetable*}

\section{Discussion}\label{sec:discussion}
We have analyzed the data of the 2013 October 11 solar flare from \fermi, \hsi,
\sdo and \stereo. \stereo-B images indicate that the flare occurred in an AR
9\de.9 behind the limb. \hsi and GBM NaI$_{1}$ detected HXRs up to 50 keV from
the flaring loop-top. The most unusual aspect of this flare is the LAT detection
of photons of energies $\epsilon >$100 MeV for about 30 minutes with some
photons having energies up to 3 GeV. Electrons or protons with energies $E >
\epsilon$ can produce these photons after traversing a column depth of matter
$N(E)>10^{25}$ and $10^{26}$ cm$^{-2}$, respectively, which is much larger than
the depth $\sim 10^{20}$ cm$^{-2}$ penetrated by HXR-producing electrons. For
occulted flares the emitted photons must traverse even larger depths where they
may be scattered and absorbed. We consider three scenarios for the emission site
of the gamma rays; (i) deep below the photosphere of the flare site (ii)  in the
corona above the limb, suggestive of trapping of particles, e.g., by strongly
converging magnetic fields and (iii) CME-shock accelerated particles traveling
back to the Sun.

\subsection{Emission below the photosphere}\label{sec:belowphotosphere}
For the first scenario we need continuous acceleration of particles
because they penetrate deep into the solar atmosphere and lose energy in a
fraction of a second. Most of the radiation they produce also comes from deep
within the photosphere so we need to calculate the optical depth,
$\tau(\e)=\sigma\times N_\g(\e)$ from  the emission site to the Earth.  For
$>$100 MeV photons the main absorption is via pair production with a cross section  $\sigma_{\rm PP}\sim 0.035\times\sigma_0$, where $\sigma_0$ is the Thomson scattering cross section relevant for $<$100 keV HXRs\footnote{For intermediate energies we are in the Klein-Nishina regime and the cross section varies smoothly between $\sigma_0$ and $\sigma_{\rm PP}$\citep{1994ApJ...434..747P}.}.
The column depth along the line of sight to the observer, $N_\g(\e)$, depends on both the position of the flare and the column depth $N(E)$ penetrated by the
emitting particles of energy $E=\eta \epsilon$. This depth is determined by the energy loss rates.

High-energy electrons spiraling down a magnetic field line with a
pitch angle cosine $\mu$ lose and radiate most of their energy deep in the
photosphere. For $E \lesssim$ 250~MeV (Lorentz factor $\gamma\lesssim 500$), the dominant energy losses are due to Coulomb-ionization, whereas for $E\gtrsim$ 250~MeV, the radiative losses are dominated by bremsstrahlung (over synchrotron and inverse Compton). The total loss rate can be approximated as

\begin{equation}\label{lossrate}
dE/dr=(1/\mu) m_ec^2(n/N_0)[1+(\gamma/\gamma_0)^\delta]
\end{equation}
with $N_0=(4\pi r_0^2\ln \Lambda)^{-1}=5\times 10^{22}\, {\rm cm}^{-2}$ (for  
Coulomb logarithm $\ln \Lambda=20$) and $n$ the total density. For extreme relativistic
electrons $\delta\sim 1.1$ and $\gamma_0$ is the Lorentz factor where the two
losses are equal. 
From these, and ignoring the small deviation of $\delta$ from unity, the column
depth penetrated by an electron of initial Lorentz factor $\gamma$ is then
\begin{equation}\label{column}
N(E) = \int_R^\infty n(r)dr = \mu N_0 \gamma_0\ln (1+\gamma/\gamma_0).
\end{equation}
For non-relativistic electrons and protons the Coulomb collision dominates and
for both particles we have
\begin{equation}\label{Coulomb}
N(E)=\mu N_0(m/m_e)(E/mc^2)^2/\gamma.
\end{equation}

The energy dependence of the proton loss rate is similar to that of electrons.
The Coulomb losses dominate at low energies but for proton energies $E>0.3$ GeV
pion production becomes significant and at $E>4.5$ GeV ($\g \gtrsim 4$) it
becomes the dominant loss mechanism, and
like electron bremsstrahlung, it gives $dE/ds\propto \g\ln \g$. Making the same
approximation as above we get the same equations, (\ref{lossrate}) and
(\ref{column}), but now with
$N_0=6\times 10^{25}$cm$^{-2}$ and $\g_0= 4.0$ (for  $\ln \Lambda=30$).

An electron of energy $E$ radiates photons of energy ${\bar \epsilon}=E/\eta <
E$ with
$\eta\sim 2$. Similarly, assuming that protons of energy $E$ produce a $\pi^0$
with similar energy that decays into two photons of equal energies we can again
set $\eta\sim 2$ (the exact value of $\eta$ will not  change our conclusions
drastically). For relativistic electrons the radiated photons will be
beamed along the pitch angle of the electrons. As shown by~\citet{1990ApJ...359..541M} there will be strong center-to-limb variation of gamma-ray
flux, but for flares at a few degrees behind the limb this effect can be
ignored.

For  a flare at the center of the solar disk
(helio-longitude $\phi=0$, or angle to the limb $\theta\equiv\pi/2-\phi=\pi/2$),
the optical depth is
$\tau(\epsilon)=N(E=\eta\epsilon)\sigma$ and  increases toward the limb at a
rate that depends on the
ambient density profile; $N_\gamma(\epsilon, \theta)=\int_0^\infty n(r)dl$,
where $ r = \sqrt{R^2 + 2Rl\sin\theta +l^2} $, and $R$
is the distance from the center of the Sun at the depth of emission $N(E)$.
The photons of interest here are produced  below the photosphere at
column depths $N>10^{25}$ cm$^{-2}$ and densities $n>10^{17}$ cm$^{-3}$, where
both quantities increase exponentially with a scale height $H \ll$ \Rs.
Consequently, most of the contribution  comes
from within a scale height at a radius $R \sim$ \Rs corresponding to the
column depth $N(\eta\epsilon)$ described above. If we define $A=R/H$ and
$d\lambda=dl/R$ we get
\begin{equation}\label{Noftheta}
N_\gamma(\epsilon, \theta)= N(E=\eta\epsilon)A e^A \int_0^\infty
e^{-A\sqrt{1+2\lambda\sin\theta+\lambda^2}}d\lambda.
\end{equation}
For occulted flares  $\theta<0$.
Since $A\gg 1$ most of the contribution to this integral comes from very small
$\lambda$ so we can use the approximation
$\sqrt{1+2\lambda\sin\theta+\lambda^2}\sim (1 + \lambda\sin\theta+ \lambda^2\cos^2(\theta)/2)$.
This gives
\begin{equation}\label{Noftheta2}
N_\gamma(\epsilon, \pm\theta)= N(\eta\epsilon)
\sqrt{\pi A/(2\cos^2\theta)}\,e^\xi\,[1 \mp
{\rm erf}(\sqrt{\xi})],
\end{equation}
where $\xi\equiv A\tan^2(\theta)/2$. Thus, we get $N_\gamma(\epsilon,
\pi/2)=N(\eta\epsilon)$ and 
$N_\gamma(\epsilon, 0)= N(\eta\epsilon)\sqrt{\pi A/2}$ 
for flares at the center
and limb of the Sun. However, we are interested
in behind-the-limb flares with $|\theta| \ll 1$  so that
$\xi\equiv\theta^2A/2=0.1(\theta^{\circ})^2(10^3 \,{\rm km/H})$. For angles
$|\theta|>3^{\circ}$ the error functions ${ \rm erf}(\sqrt{\xi})\rightarrow 1$
and $N_\gamma(\epsilon,-\theta)= 2N(\eta\epsilon)\sqrt {\pi A/2}e^{\xi}$. 

We can use these expressions to calculate the optical depth. Let us
consider HXRs emitted by nonrelativistic electrons. For a 50 keV photon
emitted by a $\mu=0.5$ and $E=100$ keV electron (and using the Thomson cross
section) we find
$\tau(50\,{\rm keV}, \theta=\pi/2)< 10^{-3}$ at the center of the disk,
$\tau(50\,{\rm keV}, \theta=0)\sim 0.02$  at the limb, and a
rapid increase as $e^{0.1\theta^2}$ for a  behind-the-limb flare. At
$\theta=-10$\de and $H=1000$ km, the optical depth is $\gtrsim 10^{3}$.

For gamma rays the optical depth is
considerably larger because they are emitted deeper in the photosphere.
For a flare near the center of the solar disk the optical depth for a 100 MeV
photon,  emitted either by a $\sim200$ MeV electron or $>350$  MeV proton, is
about 0.2 and 0.1, respectively. But for a flare at the limb these values
increase by $\sqrt{\pi A/2}= 33\times(10^3\, {\rm km/H})^{1/2}$.
For reasonable average pitch angles ${\bar \mu}>0.1$ and even if we include
the effects of non-radial field lines this could give $\tau>1$.

Once the flare source region moves behind the limb, the
optical depth increases exponentially as $e^\xi\propto e^{\theta^2A/2}$ making
the detection of any flare for $|\theta|>2$\de impossible. This is also true for
the 1 to 10 MeV BGO photons even though the protons producing them do not
penetrate as deeply. Considering that the flare occurred at $\sim$10\de behind
the limb, we conclude that this scenario is untenable.

\subsection{Emission in the corona}

The key feature of this behind-the-limb flare is the detection of $>$100 MeV emission for $\sim$30 minutes. To explain this observation, we also consider the scenario where the photons are produced in the corona by high-energy particles injected promptly into a magnetic trap~\citep[e.g.,][]{1997ApJ...487..936A} at $>10^9$ cm ( minimum height needed for a source $\sim$10$^{\circ}$ behind the limb to be visible above the limb) above the transition region.  For particles to be trapped
efficiently we need sufficiently strong field convergence to trap most
particles and a low level of turbulence 
so that the scattering time would be much longer than the energy loss time.
Otherwise most of the particles will be scattered into the loss cone and
radiate deep in
the solar atmosphere as in scenario~\ref{sec:belowphotosphere}. 

Let us first consider protons with an energy loss rate given by Eq.
(\ref{lossrate}) with $\delta=1, N_0=6\times 10^{25}$ cm$^{-2}$ and $\g_0= 4.0$.
This
gives an energy loss time for a 1 GeV proton of $\tau_0\sim 2\times 10^{15}\,
{\rm s\cdot cm}^{-3}/n$ so for the observed duration of $< 2000$ s we need a density
$n> 10^{12}\, {\rm cm}^{-3}$ which is what one encounters below the
occulted transition region and not at $> 10^9$ cm above it. The energy
loss rate for electrons in the coronal region is dominated by Coulomb
collisions at low energies and synchrotron losses
for magnetic fields $B>10$ G or inverse Compton losses at lower fields. This
rate
is again described by Eq. (\ref{lossrate}) but with $\delta=2, N_0=5\times
10^{22}$ cm$^{-2}$ and $\g_0 = 21.5 \times (n/10^{10}~{\rm cm}^{-3})^{1/2}
\times (B/100~{\rm G})^{-1}$. As shown in \citet{2001ApJ...557..560P}, this
gives rise to flat spectra at low energies and a sharp cutoff at $\g\sim\g_0$
with these electrons carrying most of the energy with the longest lifetime of
$\tau_0\sim \g_0N_0/(2nc)$. For the production of $>100$~MeV photons we
need electrons with $\g_0\sim 300$ so that for a lifetime of 30 minutes we need $n\sim 10^{11}\,{\rm cm}^{-3}$ and $B\sim$ 25 G. While these values for the density (magnetic field) are somewhat higher (lower) than the ones found at 10$^{9}$ cm above the transition region, they cannot be fully ruled out. Also, a photon index $\Gamma\sim 0$ requires injected
electrons with spectral index  -1 (for a thin target case), which is much harder
than those encountered at lower energies. Thus, this model requires strong
convergence, low turbulence and hard spectra. In view of the LAT detection of SOL2014-09-01 flare with $\theta\sim 36^{\circ}$
(paper in preparation) that requires a trap at a height $>$10$^{10}$ cm, this
model becomes less plausible.

\subsection{Acceleration in CME Shocks}

A third possibility is that particles accelerated by a shock associated with
this flare originating behind the limb find their way to the photosphere visible
to \Fermi where they produce gamma-rays. This requires a magnetic connection
between the acceleration site and the visible photosphere, e.g., large overlying
loops. Such a connection must have been absent during the impulsive phase and
for HXR-producing electrons which are most likely accelerated in smaller loops
with both footpoints occulted. Otherwise we would expect a detection of HXRs
from the footpoints on the visible side. Furthermore, the LAT emission error
circle allows for the gamma-ray emission to occur on the visible side of the
disk. This also means that the extended-phase gamma-ray producing particles were
not accelerated in small loops and, like longer lasting SEPs, most likely were
accelerated in the shock of the CME. Since the magnetic lines draping the CME
are most likely connected to the occulted AR, this requires cross-field
diffusion that allows migration of particles from the field lines connected to
the AR to those connected to the visible disk. The presence of a strong and
short scale turbulence capable of scattering the accelerated particles with a
mean free path comparable to their gyro radii will facilitate this migration
\citep{2003ApJ...595..493Z}. A longer trapping time of accelerated
particles in the downstream region, say e.g., by converging field lines rooted at the Sun, can also help this migration. This model requires further study.

In conclusion, the multiwavelength observations of this behind-the-limb flare
have provided some interesting theoretical puzzles which can be resolved by more
detailed investigation of the scenarios discussed above.

\acknowledgements
The $Fermi$ LAT Collaboration acknowledges support from a number of agencies and
institutes for both development and the operation of the LAT as well as
scientific data analysis. These include NASA and DOE in the United States,
CEA/Irfu and IN2P3/CNRS in France, ASI and INFN in Italy, MEXT, KEK, and JAXA in
Japan, and the K.~A.~Wallenberg Foundation, the Swedish Research Council and the
National Space Board in Sweden. Additional support from INAF in Italy and CNES
in France for science analysis during the operations phase is also gratefully
acknowledged.
V.P, W.L and F.R.d.C are supported by NASA grants NNX14AG03G, NNX13AF79G and
NNX12AO70G. The authors thank Eric Grove and Ron Murphy for helpful
suggestions.
%
%

\end{document}